\documentclass[preprintnumbers,aps,prl,twocolumn,superscriptaddress,notitlepage,footinbib]{revtex4-1}
\usepackage[english]{babel}
\usepackage[utf8]{inputenc}
\usepackage[colorinlistoftodos, color=green!40, prependcaption]{todonotes}

\usepackage{amsthm}
\usepackage{mathtools}
\usepackage{physics}
\usepackage{xcolor}
\usepackage{graphicx}
\usepackage{adjustbox}
\usepackage{placeins}
\usepackage[T1]{fontenc}
\usepackage{lipsum}
\usepackage{csquotes}

\usepackage{lineno}
\modulolinenumbers[5]

\usepackage[colorlinks=true, urlcolor=blue,citecolor=blue,anchorcolor=blue]{hyperref}

\newcommand\pT{\ensuremath{p_\mathrm{T}}}

\newcommand{\sectionPRL}[1]{ \textbf{ #1.}}

\newcommand{\jet} {\text{jet}}

\newcommand{\XXXX}{Centauro}   
\usepackage{ulem}

\begin{document}

\preprint{JLAB-THY-20-3209}

\title{Asymmetric jet clustering in deep-inelastic scattering}

\author{M.~Arratia}
\email{miguel.arratia@ucr.edu}
\affiliation{Physics Department, University of California, Riverside, CA 92521, USA}
\affiliation{Thomas Jefferson National Accelerator Facility, Newport News, VA 23606, USA}

\author{Y.~Makris}
\email{yiannis.makris@pv.infn.it}
\affiliation{INFN Sezione di Pavia, via Bassi 6, I-27100 Pavia, Italy}

\author{D.~Neill}
\email{duff.neill@gmail.com}
\affiliation{Theoretical Division, MS B283, Los Alamos National Laboratory, Los Alamos, NM 87545, USA}

\author{F.~Ringer}
\email{fmringer@lbl.gov}
\affiliation{Nuclear Science Division, Lawrence Berkeley National Laboratory, Berkeley, California 94720, USA}
\affiliation{Physics Department, University of California, Berkeley, CA 94720, USA}

\author{N.~Sato}
\email{nsato@jlab.org}
\affiliation{Theory Center, Jefferson Laboratory, Newport News, Virginia 23606, USA}

\date{\today} 

\begin{abstract}
We propose a new jet algorithm for deep-inelastic scattering (DIS) that accounts for the forward-backward asymmetry in the Breit frame. The \XXXX~algorithm is longitudinally invariant and can cluster jets with near-to Born kinematics, which enables novel studies of transverse-momentum-dependent observables. Furthermore, we show that spherically-invariant algorithms in the Breit frame give access to low-energy jets from current fragmentation. We perform a calculation of the energy spectrum of Centauro jets at next-to-leading logarithmic accuracy and compare to Pythia simulations. We furthermore propose novel studies in unpolarized, polarized, and nuclear DIS at the future Electron-Ion Collider.
\end{abstract}

\maketitle
\sectionPRL{Introduction} Understanding the structure of nucleons and nuclei in terms of quark and gluons remains an open goal. Jet production in deep inelastic scattering (DIS) provides an excellent tool for this endeavor. The future Electron-Ion Collider (EIC)~\cite{Accardi:2012qut} will produce the first jets in polarized and nuclear DIS, which will enable a rich jet program~\cite{Arratia:2020azl,Borsa:2020ulb,Peccini:2020tpj,Guzey:2020gkk,Guzey:2020zza,Kang:2020xyq,Arratia:2019vju,Page:2019gbf,Li:2020sru,Gutierrez-Reyes:2019vbx,Gutierrez-Reyes:2019msa,Gutierrez-Reyes:2018qez,Zhang:2019toi,Aschenauer:2019uex,Hatta:2019ixj,Mantysaari:2019csc,DAlesio:2019qpk,Kishore:2019fzb,Kang:2019bpl,Roy:2019hwr,Salazar:2019ncp,Boughezal:2018azh,Klasen:2018gtb,Dumitru:2018kuw,Liu:2018trl,Zheng:2018ssm,Sievert:2018imd,Klasen:2017kwb,Hinderer:2017ntk,Chu:2017mnm,Abelof:2016pby,Hatta:2016dxp,Dumitru:2016jku,Boer:2016fqd,Dumitru:2015gaa,Hinderer:2015hra,Altinoluk:2015dpi,Kang:2013nha,Pisano:2013cya,Kang:2012zr,Kang:2011jw,Boer:2010zf}.  

The HERA jet measurements in DIS targeted gluon-initiated processes by requiring large transverse momentum in the Breit frame~\cite{Newman:2013ada}. This suppresses the Born configuration, $\gamma^{*}q\to q$, which has recently been postulated as key to probe transverse-momentum dependent (TMD) PDFs~\cite{Gutierrez-Reyes:2018qez,Gutierrez-Reyes:2019vbx,Gutierrez-Reyes:2019msa}. Complementary to semi-inclusive DIS observables, jets avoid nonperturbative TMD fragmentation functions. Moreover, modern jet substructure techniques~\cite{Larkoski:2017jix} offer
new methods for precise QCD calculations and to control nonperturbative effects, e.g grooming or a recoil-free axis can be used to minimize hadronization effects or study TMD evolution~\cite{Makris:2017arq}. These techniques also provide new ways to connect to lattice QCD calculations, e.g. of the nonperturbative Collins-Soper kernel~\cite{Ebert:2018gzl,Shanahan:2020zxr}. 

\begin{figure}
  \centerline{\includegraphics[width = 0.42 \textwidth]{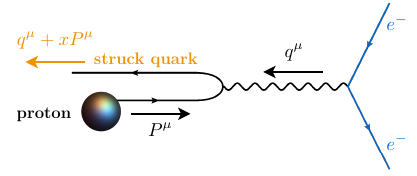}}
  \caption{DIS Born kinematics in the Breit frame. }
  \label{fig:diagram}
\end{figure}

The Breit frame plays a central role in jet clustering for DIS~\cite{Webber:1993bm}, and it allows for a factorized TMD cross-section in terms of the same soft and un-subtracted TMD functions as in Drell-Yan and $e^+e^- \to$ dihadron/dijet processes~\cite{Gutierrez-Reyes:2018qez,Gutierrez-Reyes:2019vbx,Gutierrez-Reyes:2019msa}. However, the longitudinally-invariant (LI) algorithms commonly used in DIS cannot cluster jets that enclose the beam axis given by the proton/photon direction (see Fig.~\ref{fig:diagram}). 

In this letter, we introduce a new jet algorithm that is longitudinally invariant but can capture jets close to the Born configuration in the Breit frame. In addition, we use spherically-invariant (SI) algorithms to study the jet energy spectrum and find that they can separate the current and target fragmentation region even for soft jets. Finally, we suggest novel studies of jet energy and TMD observables.  

\begin{figure*}
  \centerline{\includegraphics[width= 0.98 \textwidth]{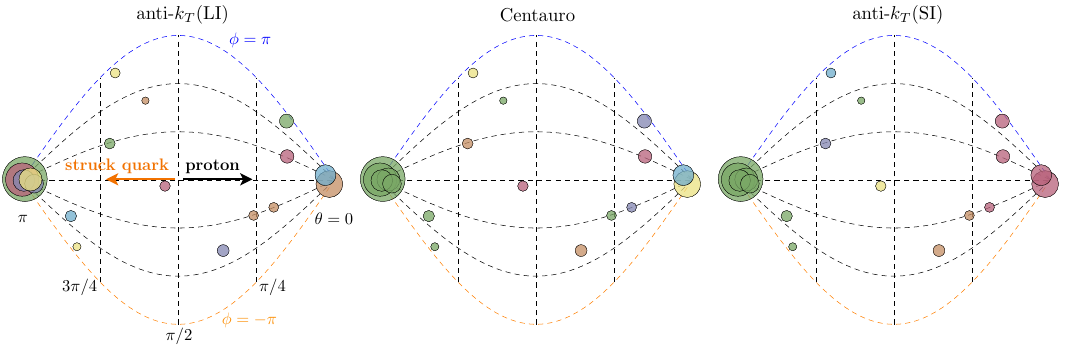}}
  \caption{Jet clustering in the Breit frame using the longitudinally-invariant anti-$k_{T} (\text{LI})$, \XXXX, and spherically-invariant anti-$k_T (\text{SI})$ algorithms in a DIS event simulated with \textsc{Pythia} 8. Each particle is illustrated as a disk with area proportional to its energy and the position corresponds to the direction of its momentum projected onto the unfolded sphere about the hard-scattering vertex. The vertical dashed lines correspond to constant $\theta$ and curved lines to constant $\phi$. All the particles clustered into a given jet are colored the same.}
  \label{fig:demo}
\end{figure*}

\sectionPRL{Notation and DIS kinematics} In the Breit frame, the virtual photon momentum is given by:
\begin{equation}
    q^{\mu}  = \frac{Q}{2}( \bar{n}^{\mu} - n^{\mu} ) = Q (0,0,0,-1)\,,
\end{equation}
where $n^{\mu} \equiv (1,0,0,+1)$ and $\bar{n}^{\mu} \equiv (1,0,0,-1)$.
The proton momentum (up to mass corrections) is:
\begin{equation}
    P^{\mu} \simeq Q/(2x_B) n^{\mu} = Q/(2x_B) (1,0,0,+1)\,,
\end{equation} 
with Bjorken $x_B \equiv Q^2 / (2 \,q\cdot P)$. At Born level, the struck quark back-scatters against the proton and has momentum ($x\simeq x_B$):
\begin{equation}
    p_q^{\mu} = x P^{\mu} + q^{\mu} \simeq (Q/2) \bar{n}^{\mu}\,.
\end{equation}
The fragmentation of the struck-quark yields a jet that points to the beam direction. The algorithms we introduce below are designed to capture this jet. We define the scaling variable: 
\begin{equation}
    z_\jet = \frac{P \cdot p_\jet}{P \cdot q} \;\;\;\xrightarrow[\text{frame}]{\text{Breit}} \;\;\; z_\jet = n\cdot p_{\rm jet}/Q= p^+_\jet / Q \,.
\end{equation}
At leading-logarithmic accuracy, $z_\jet$ is the fraction of the struck-quark momentum carried by the jet. 

 
\sectionPRL{New jet algorithms for DIS}
The longitudinally-invariant $k_T$-type jet algorithms~\cite{Ellis:1993tq, Dokshitzer:1997in, Wobisch:1998wt, Cacciari:2008gp,Cacciari:2011ma} use the following distance measure: 
\begin{align}
    \label{eq:anti-kT-LI}
    d_{ij}  &= \min(p_{Ti}^{2p},p_{Tj}^{2p}) \Delta R^2_{ij}/R^2\;,&d_{iB} =  p_{Ti}^{2p}\,,
\end{align}
where $\Delta R_{ij} =  (y_i -y_j)^2 +  (\phi_i -\phi_j)^2$. Here $d_{ij}$ is the distance between two particles in the event and $d_{iB}$ is the beam distance.
Since they cluster particles in the rapidity-azimuth ($y -\phi$) plane, they cannot form a jet enclosing the $\bar{n}^{\mu}$ direction $(y=-\infty)$.

\begin{figure*}
  \centerline{\includegraphics[width = 0.99 \textwidth]{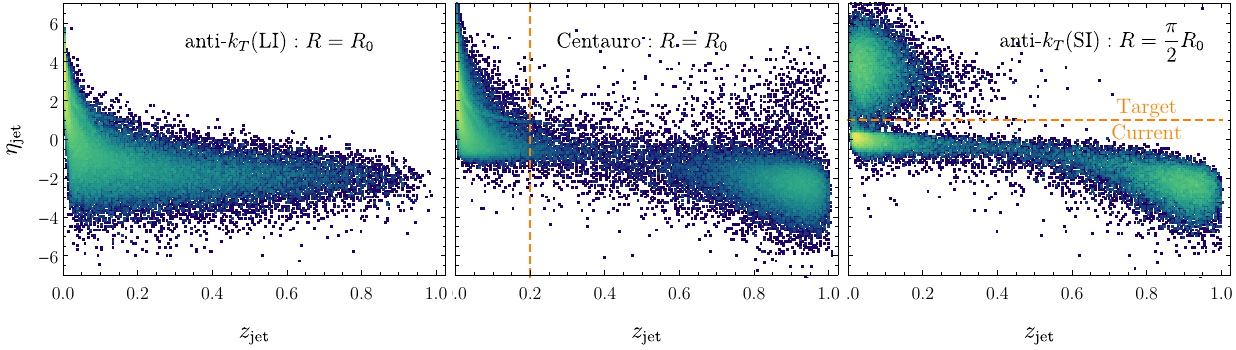}}
  \caption{The distribution of jets in the Breit frame in terms of their pseudorapidity $\eta_\jet$ and momentum fraction $z_\jet$. The left, center, and right panels correspond to jets identified with the anti-$k_T(\text{LI})$, \XXXX, and anti-$k_T(\text{SI})$ algorithms, respectively. The dashed lines indicate the separation of jets in the current and target fragmentation region.}
  \label{fig:scatterETAvsZ}
\end{figure*}

One way to bypass this problem is to use spherically-invariant algorithms. Catani et al. first proposed to adapt spherically-invariant algorithms to DIS in ref.~\cite{Catani:1992zp}. In this study  we consider the $k_T$-type algorithms for $e^+e^-$ collisions~\cite{Catani:1991hj, Cacciari:2011ma}, which have the following distance measure:
\begin{align}
    \label{eq:anti-kT-SI}
    d_{ij}  &= \min(E_{i}^{2p},E_{j}^{2p})  \frac{1 - c_{ij}}{1-c_R} \;,&d_{iB} =  E_{i}^{2p} \,,
\end{align}
where $c_{ij} = \cos\theta_{ij}$ and $c_R=\cos R$. However, these algorithms lack the longitudinal invariance that connects the class of frames related to the Breit frame by $\hat{z}$ boosts, which is a crucial feature of jet clustering~\cite{Webber:1993bm}. For example, it is important for multijet events where the parton kinematics is not constrained by $x_B$ and $Q^{2}$, and to identify photo-production or separate the beam remnant from forward jets~\cite{Catani:1993hr}. 

To solve this issue, we introduce a new jet algorithm that is longitudinally invariant along the Breit frame beam axis but yet captures the struck-quark jet. Recently, Boronat et al.~\cite{Boronat:2014hva} proposed a hybrid algorithm that suppresses $\gamma\gamma$ background in $e^{+}e^{-}$ colliders. In contrast, we suggest a jet algorithm that is asymmetric in the backward and forward directions, and suggest novel studies for spherically-invariant algorithms in DIS. 

Starting with the distance measure of the Cambridge/Aachen (C/A) algorithm for $e^+e^-$ (i.e Eq.~(\ref{eq:anti-kT-SI}) for $p=0$),
we write the numerator in Eq.~(\ref{eq:anti-kT-SI}) in terms of the unit vectors along the directions of particles $i$ and $j$,
\begin{align}
    1 - c_{ij} &= 1- \hat{n}_i \cdot \hat{n}_j = 1 - s_i s_j \cos\Delta\phi_{ij} - c_i c_j\,.
\end{align}
with $c_i = \cos\theta_i$ and $s_i = \sin\theta_i$. Expanding in the very backward limit (i.e. $\bar{\theta}_i \equiv \pi -\theta_i \ll 1$) we find:
\begin{equation}\label{eq:rot-app}
    1-c_{ij} \simeq \frac{1}{2}(\bar{\theta}_i -\bar{\theta}_j)^2 + \bar{\theta}_i \bar{\theta}_j (1-\cos\Delta \phi_{ij}).
\end{equation}
We then introduce the replacements: 
\begin{align}
\bar{\theta}_i & \to f_i = f ( \bar{\eta}_i )\,, & 
\bar{\eta}_i &\equiv  -  \frac{2 Q}{\bar{n}\cdot q} \frac{ p_i^{\perp}} {   n \cdot p_i }\,,
\end{align}
where the function $f$ must satisfy: $f (x) = x +\mathcal{O}(x^2)$, and $p^{\perp}_i$ is the transverse momentum in the Breit frame. The term $\bar{\eta}_i$ (which in the Breit frame is $ 2 p^{\perp}_i /(n \cdot p_i) $) introduces an asymmetry: in the backward region the distance between particles is given by their separation in $\bar{\eta}$, which decreases as particles become closer in angle. In contrast, in the forward region $\bar{\eta}$ diverges
and thus prevents jets from enclosing the proton beam direction, like the anti-$k_T\text{(LI)}$ algorithm. We thus introduce the following distance measure:
\begin{align}
    d_{ij} &= \Big{[}(\Delta f_{ij})^2 +2 f_i f_j (1-\cos \Delta \phi_{ij})\Big{]} / R^2\;, & d_{iB} = 1
\end{align}
which defines a new class of algorithms, which we call \XXXX~algorithms. Two relevant choices~\footnote{Different choices can be optimized to match jets found by other algorithms, for example we found that jets clustered in the forward region are better matched to the jets clustered with the anti$-k_{T}$ (LI) algorithm.} for the function $f$ are:
\begin{align}
    f(x) & = x\,, &f(x)& = \sinh^{-1}(x)\,.
\end{align} 
The Centauro algorithm is invariant along the $\hat{z}$ direction, but in the backward hemisphere it matches the spherically-invariant algorithms (see Eq.~(\ref{eq:anti-kT-SI})). This feature is largely independent of the choice of $f$~\footnote{For the analysis in this section we started with the choice $p=0$ in Eq.(\ref{eq:anti-kT-SI}). Analogous algorithms can be obtained for the choices $p=+1$ and $p=-1$.}.

\sectionPRL{Simulation results and applications} Throughout this letter we analyze DIS events with $Q > 10$~GeV simulated in \textsc{Pythia}~8~\cite{Sjostrand:2007gs} 
with 10 and 100 GeV electron and proton beam energies respectively
~\footnote{In addition, we verified that our conclusions are the same when using the \textsc{DIRE} shower of ref.~\cite{Hoche:2015sya}.}. We exclude neutrinos and particles with $\vert \eta \vert > 4$ or $p_T < 200$ MeV in the laboratory frame. We use~\textsc{Fastjet}~\cite{Cacciari:2011ma} to cluster jets in the laboratory frame with the anti-$k_T (\text{LI})$, anti-$k_T (\text{SI})$, and Centauro algorithms~\footnote{The Centauro algorithm is available as part of the official release of the Fastjet Contrib package, since fjcontrib-1.045, see~\url{https://fastjet.hepforge.org/contrib/}.}; Fig.~\ref{fig:demo} illustrates the resulting jet clustering for an exemplary \textsc{Pythia}~8 event. The anti-$k_T (\text{LI})$ algorithm clusters the particles from the fragmentation of the struck-quark into four different jets~\footnote{Note that this cannot be corrected for by choosing a larger jet radius due to the exponential increase of the rapidity in the forward and backward region.}. In contrast, the anti-$k_T (\text{SI})$ and \XXXX~algorithms cluster all of these particles into a single jet with $z_\jet \sim 1$. The \XXXX~algorithm mimics the features of the anti-$k_T (\text{SI})$ in the backward direction and the anti-$k_T (\text{LI})$ in the forward direction.

Furthermore, with the use of \XXXX~and anti-$k_T(\text{SI})$ jets it is also possible to suppress the target fragmentation with a cut on $z_\jet\sim 0.2-0.7$, as shown in Fig.~\ref{fig:scatterETAvsZ} (center panel). This allows for direct studies of quark TMD observables. For the anti-$k_T(\text{SI})$~\footnote{Note that we include a factor of $\pi/2$ for the jet radius of the anti-$k_T(\text{SI})$ algorithm to account for the difference in the denominator of the clustering metric in eq.~(\ref{eq:anti-kT-SI}) compared to eq.~(\ref{eq:anti-kT-LI}).} algorithm, a cut on $\eta_\jet < 1$ separates current and target regions (right panel of Fig.~\ref{fig:scatterETAvsZ}). This reveals the full $z_\jet$ spectrum, which cannot be accessed with hadron measurements due to the contamination from the target fragmentation region~\cite{Boglione:2016bph,Aschenauer:2019kzf}.  For comparison we also show the result for the anti-$k_T(\text{LI})$ algorithm in the left panel of Fig.~\ref{fig:scatterETAvsZ}.

\begin{figure}
  \centerline{\includegraphics[width = 0.42 \textwidth]{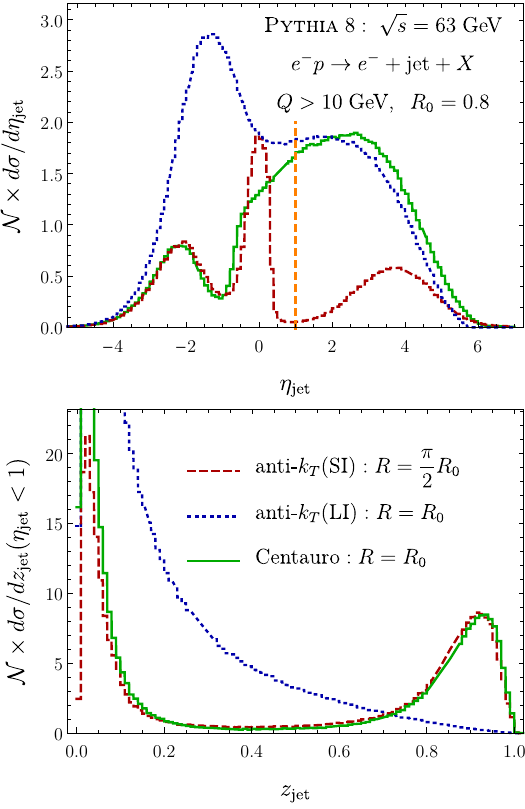}}
          \caption{Pseudorapidity (top panel) and momentum fraction $z_\jet$ (bottom panel) of jets clustered with anti-$k_{T}(\text{LI})$, anti-$k_T(\text{LI})$ and \XXXX~algorithms in the Breit frame. Here $\mathcal{N}$ is an overall normalization constant chosen to improve readability and is the same for all curves in a graph.}
  \label{fig:LFvsBF}
\end{figure}

Fig.~\ref{fig:LFvsBF} shows the $z_\jet$ and $\eta_\jet$ distributions of inclusive jets as described above. While in the backward region ($\eta_\jet < 0$), the \XXXX~and anti-$k_T(\text{SI})$ algorithms result in a peak at large $z_\jet \sim 1$, the anti-$k_T(\text{LI})$ algorithm separates that jet into several and yields a peak at small-$z_\jet$. The two peaks at $z_\jet \sim 1$ and $z_\jet \sim 0$ correspond to backward and mid rapidity jets. The intermediate $z_\jet$ region is described in terms of jet functions and DGLAP evolution~\cite{Dasgupta:2014yra,Kaufmann:2015hma,Kang:2016mcy,Dai:2016hzf}. The large-$z_\jet$ jets probe the threshold region~\cite{Dai:2017dpc}, whereas the small-$z_\jet$ region is related to soft fragmentation in $e^+e^-$ collisions~\cite{Mueller:1981ex,Bassetto:1982ma,Dokshitzer:2005bf,Kom:2012hd,Anderle:2016czy} and small-$x$ physics~\cite{Basso:2006nk,Hatta:2008st,Neill:2020bwv}.
\begin{figure}
  \centerline{\includegraphics[width = 0.42 \textwidth]{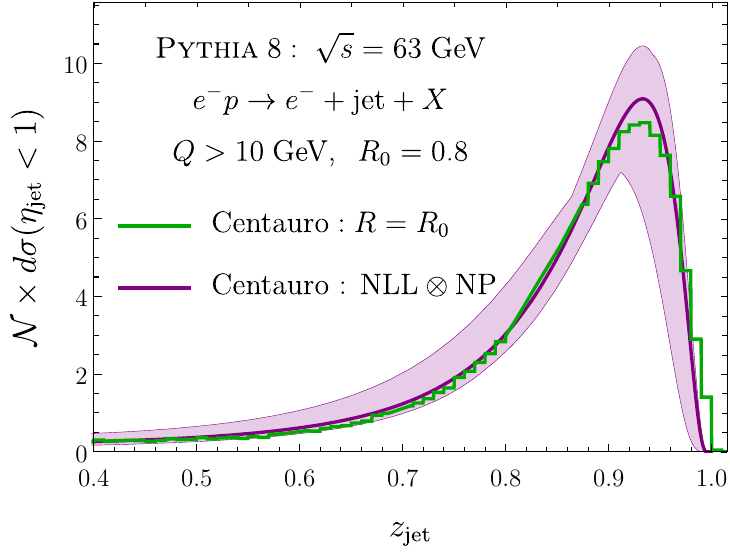}}
  \caption{The NLL prediction for the Centauro algorithm including threshold effects to NLL accuracy, full DGLAP to LL, as well as a nonperturbative shape function. }
  \label{fig:theory}
\end{figure}
 In Fig.~\ref{fig:theory}, we show also the energy spectrum for jets that results from a perturbative calculation supplemented with a nonperturbative shape function. As detailed in the Appendix, the spectrum results from the calculation of the factorization formula:
{\small\begin{align*}
d\sigma=\sigma_0 H(Q^2,\mu^2)\!\!\!\int\displaylimits_{z_{\rm jet}}^{1}\!\!\frac{dz}{z}\!B_{q}\Big(\!x_B,Q^2(1\!-\!z),\mu^2\!\Big)D_{q}\Big(\!\frac{z_{\rm jet}}{z},QR,\mu^2\!\Big),
\end{align*}}
where the formula is differential in $x_B,Q^2$ and $z_{\rm jet}$. The function $B_q$ is the quark beam function of refs. \cite{Fleming:2006cd,Stewart:2009yx} and $D_q$ is the quark fragmentation function to a jet at the endpoint from ref.~\cite{Dai:2017dpc}. The resummation formula at the end-point can be derived by combining the methods developed in refs. \cite{Fleming:2006cd,Kang:2016mcy,Dai:2016hzf,Dai:2017dpc,Liu:2017pbb,Lustermans:2019cau}, valid to next-to-leading logarithm (NLL) including non-global effects of refs.~\cite{Dasgupta:2001sh,Banfi:2002hw}. We also matched to the full leading order DGLAP evolution in the moderate $z_{\rm jet}$ region. Exploiting the sum-rule for the jets which demands conservation of the final state momentum, we can normalize to the leading order DIS cross-section. The PDFs were obtained from refs.~\cite{Clark:2016jgm}. The NLL uncertainty band is obtained from the envelope of varying each low scale of the renormalization group evolution by a factor of two, as well as all nonperturbative shape function scales and cutoffs for the Landau pole.

We propose a measurement of $z_\jet$ at HERA, which has not been done before, and the future EIC. The high-$z_\jet$ region corresponds to jets with high-\pT~in the laboratory frame that can be measured with high precision and with an accuracy limited by the jet energy scale uncertainty, which reached 1$\%$ at HERA~\cite{Newman:2013ada}. The measurement of the small-$z_\jet$ region will be challenging because these jets correspond to jet \pT~up to a few GeV in the laboratory frame~\footnote{This depends on $Q$, and a high $Q$ is preferred to reach the lowest $z_\jet$}, a region that can be limited by calorimeter noise and resolution. These issues could be bypassed by defining jets with charged particles only, which would require the inclusion of track-based jet functions on the theory side~\cite{Chang:2013iba, Chang:2013rca}. 

\begin{figure}
  \centerline{\includegraphics[width = 0.42 \textwidth]{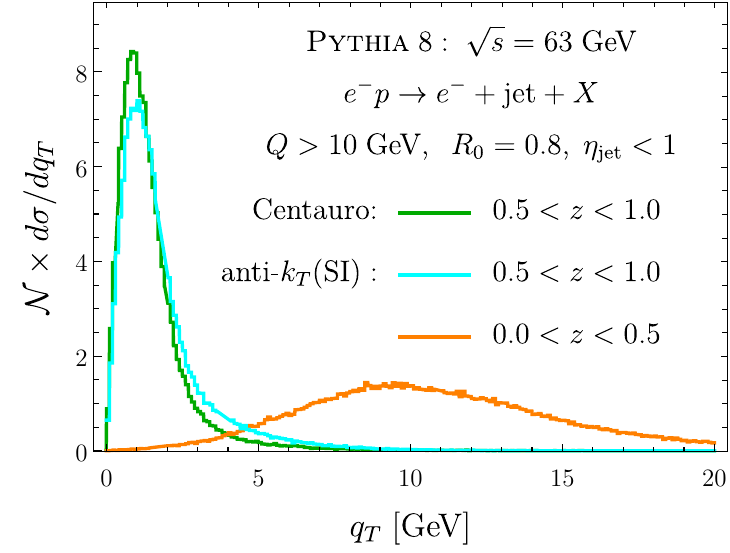}}
  \caption{The $q_T = p_\jet^{\perp} /z_\jet$ spectrum for \XXXX~and anti-$k_T(\text{SI})$ jets with $\eta_\jet <1.0$.}
  \label{fig:TMD}
\end{figure}

We also propose to use \XXXX~jets to study quark TMDs by measuring $q_T = p_\jet^{\perp} /z_\jet$. Fig.~\ref{fig:TMD} shows that the $q_T$ spectrum for $z_\jet >0.5$ peaks at $q_T < Q_{\text{min}} /4$, which is ideal for TMD phenomenology. With the polarized proton beams available at the EIC, this observable would provide clean access to the Sivers PDFs. 
Fig.~\ref{fig:TMD} also shows the $q_T$ distribution for anti-$k_T(\text{SI})$ jets for $z_\jet > 0.5$ and  $ 0 < z_\jet < 0.5$. Note the latter is only possible since we can suppress the target fragmentation by requiring $\eta_\jet < 1$. While for  $z_\jet > 0.5$ we find similar result as the \XXXX~jets, for $z_\jet < 0.5$ the spectrum peaks at $q_T \sim Q$. Novel theoretical techniques are necessary to describe this kinematic $q_T$ region of mid-rapidity jets.

In addition, the longitudinal invariance and ability to measure jets close to Born kinematics makes the \XXXX~algorithm an attractive option to: i) extract the strong-coupling constant from the rates of $n$-jets~\cite{Newman:2013ada}; ii) enable ``tag-and-probe'' studies of nuclei~\cite{Arratia:2019vju}; iii) identify the background for gluon helicity and Sivers PDF studies~\cite{Page:2019gbf,Zheng:2018ssm}, iv) study jet substructure and event shape observables in DIS Breit frame, v) probe TMD evolution observables that can be related to lattice QCD.  We leave those studies for future work. 

\sectionPRL{Conclusions} We have proposed a new jet clustering approach tailored to the study of energetic jets with low transverse momentum in DIS that relies on spherically-invariant algorithms and a new longitudinally-invariant algorithm that is asymmetric in the backward and forward directions, which we call \XXXX. The \XXXX~algorithm enables novel studies of transverse-momentum-dependent observables in the Breit frame. Furthermore, we find that spherically symmetric $k_{T}$-type algorithms yield clean access to the soft jet fragmentation region, which also reveals a new $q_T$ regime where $q_T\sim Q$. The new jet algorithms introduced here are relevant for the studies of jet energy spectra, jet substructure, quark TMDs and spin physics, and cold-nuclear matter effects. All these studies will be central for the jet physics program of the future Electron-Ion Collider.

\begin{acknowledgments}
\sectionPRL{Acknowledgements} 
We thank Anselm Vossen, Renee Fatemi, and Fernando Torales-Acosta and 
Benjamin Nachman for insightful comments on our manuscript.  M.A and N.S are supported through DOE Contract No. DE-AC05-06OR23177 under which JSA operates the Thomas Jefferson National Accelerator Facility. Y.M is supported by the European Union’s Horizon 2020 research and innovation programme under the Marie Skłodowska-Curie grant agreement No. 754496 - FELLINI. D.N. was supported by the U.S. DOE under Contract DE-AC52-06NA25396 at LANL and through the LANL/LDRD Program. F.R. was supported by LDRD funding from Berkeley Lab provided by the U.S. Department of Energy under Contract No. DE-AC02-05CH11231 as well as the National Science Foundation under Grant No. ACI-1550228.
\end{acknowledgments}

\bibliographystyle{h-physrev}
\bibliography{bibliography.bib}

\widetext
\clearpage

\appendix
\section{Factorization and threshold resummation~\label{sec:appB}}

In this appendix, we give the necessary technical details behind the theory prediction for the jet energy spectrum using the Centauro algorithm. We work to next-to-leading logarithmic order, where we resum all logarithms found in the cross-section of order $\alpha_s\times L$, with $L=$ln$(1-z_{\rm jet})$ or ln$\,R$, $R$ the jet radius and $z_{\rm jet}$ is the momentum fraction of the event carried by the jet. This is includes logarithms of the jet radius outside the threshold limit where $1-z_{\rm jet}\ll 1$, which is formally a leading logarithmic resummation within strictly collinear factorization. The relevant details for the theory of semi-inclusive jet production and the jet function's threshold factorization can be found in refs. \cite{Kang:2016mcy,Dai:2016hzf,Dai:2017dpc,Liu:2017pbb}, while the necessary details for the treat of the initial state for the threshold jet can be adapted from ref. \cite{Fleming:2006cd,Lustermans:2019cau}. See also refs.~\cite{Cacciari:2001cw,Anderle:2012rq}. For the reader unfamiliar with the technology of resummation via soft-collinear effective field theory, we suggest ref.~\cite{Becher:2014oda}. The resummation of the jet spectrum is accomplished via the factorization formulas:
\begin{align}
\label{eq:mod_z_fact}d\sigma&=\tilde{\sigma}_0 \!\sum_{i,j}\int\displaylimits_{x_B}^{1}\!\!\frac{dx}{x}\!\int\displaylimits_{z_{\rm jet}}^{1}\!\!\frac{dz}{z}\,C_{ij}\Big(x,z,Q^2,\mu^2\!\Big)f_{i/P}\!\Big(\frac{x_B}{x},\mu^2\!\Big)D_{j}\Big(\!\frac{z_{\rm jet}}{z},QR,\mu^2\!\Big),\\
\label{eq:large_z_fact}d\sigma&=\sigma_0 H(Q^2,\mu^2)\!\!\!\int\displaylimits_{z_{\rm jet}}^{1}\!\!\frac{dz}{z}\!B_{q}\Big(\!x_B,Q^2(1\!-\!z),\mu^2\!\Big)D_{q}\Big(\!\frac{z_{\rm jet}}{z},QR,\mu^2\!\Big).
\end{align}
The sum is over the flavor indices of QCD, and $C$ is the matching coefficient for collinear factorization in semi-inclusive DIS, $H$ is the current matching squared for DIS processes, $f$ is the parton distribution function, and $D,B$ are the quark or anti-quark fragmentation function to a jet and the inclusive beam function. Eq. \eqref{eq:mod_z_fact} is valid when $1-z_{\rm jet}\sim 1$, while Eq. \eqref{eq:large_z_fact} controls the region $1-z_{\rm jet}\ll 1$. Finally,  $\tilde{\sigma}_0$ and $\sigma_0$ are the born-level cross-sections for each factorization. The two formulas are related via the operator product expansion for the beam function:
\begin{align}
 B_{i}\Big(\!x_B,Q^2(1\!-\!z),\mu^2\!\Big)&=\sum_{j}\int\displaylimits_{x_B}^{1}\frac{dx}{x}\mathcal{I}_{ij}\Big(x,Q^2(1\!-\!z),\mu^2\!\Big)f_{j/P}\Big(\frac{x_B}{x},\mu^2\Big)+\mathcal{O}\Big(\frac{\Lambda^2_{QCD}}{Q^2(1\!-\!z)}\Big)\,,
\end{align}
with $\Lambda_{QCD}$ the scale of confinement. Moreover, $D$ receives its own factorization in the endpoint region. Running each function to its natural scale (for detailed discussion, see ref. \cite{Almeida:2014uva}), evaluating them at their tree-level expressions, and factoring the PDFs, while using a nonperturbative shape function gives: 
\begin{align}
d\sigma&=\sigma_0\Big(\sum_{q'}f_{q'/P}(x_B,\mu_F^2)\Big)\Bigg(\sum_{j}\int\displaylimits_{z_{\rm jet}}^{1} \frac{dz}{z}U_{qj}(z_{\rm jet}/z;\mu^2_H,\mu^2_J)\Big[\mathcal{S}_j\otimes\frac{d\mathcal{R}_{j}}{dz}\Big]\Big(z,QR;\mu_H^2,\mu_J^2,\mu_F^2,\mu_{cs}^2\Big)\Bigg)\,,\\
    \mu_{F}^2&\sim Q^2(1-z_{\rm jet})\,,\qquad  \mu_H^2\sim Q^2\,,\qquad
    \mu_J^2\sim Q^2R^2\,,\qquad  \mu_{cs}^2\sim Q^2R^2(1-z_{\rm jet})^2\,.
\end{align}
 We factorize the beam function from the PDFs at the beam function scale, which we take as $\mu_B^2=\mu_{F}^2\sim Q^2(1-z_{\rm jet})$. $\mathcal{R}_j$ is the NLL radiator function (described below for the case of quark jets), and $U_{ij}$ is the \emph{full} NLO DGALP evolution evolved from the scale $\mu_H$ to the jet scale $\mu_J$. This is so that the formula is valid to leading log when $z_{\rm jet}\sim 0.5$, away from the end point, but will have the two loop cusp at the endpoint, so that it is still $NLL$ valid at the endpoint. For NLL, several simplifications occur: note that at the factorization scale for the PDFs, we only probe the quark and anti-quark content of the proton, so we restrict the PDF flavor sum accordingly. Moreover, the initial hard parton that generates the jets will also be a quark or anti-quark, and since these jets have identical jet functions, this restricts us to the singlet sector of the DGLAP evolution.  $\mathcal{S}_j$ is a nonperturbative shape function which we define as:
 \begin{align}
    \Big[\mathcal{S}_j\otimes\frac{d\mathcal{R}_{j}}{dz}\Big]\Big(z,QR;\mu_H^2,\mu_J^2,\mu_F^2,\mu_{cs}^2\Big)&=\int_{z}^{1}\frac{dz'}{z'} \mathcal{S}(z',d_j)\frac{d\mathcal{R}_{j}}{dz}\Big(\frac{z}{z'},QR;\mu_H^2,\mu_J^2,\mu_F^2,\mu_{cs}^2\Big)\,,\\
    \mathcal{S}(z,d)&=\frac{(1-z)\text{exp}\Big(-\frac{1}{d}(1-z)\Big)}{d(d-(1+d)e^{-\frac{1}{d}})}\,,\qquad
    d_j\sim\frac{\Lambda_{j}}{QR}\,.
\end{align}
$\Lambda_j\sim 400$~MeV$\sim\Lambda_{QCD}$ should be of the order of the confinement scale, and we have normalized the integral of $S$ to be one when integrated between zero and one. Note that we allow $\Lambda_{j}$ to be different for quark or gluon jets, but in practice we take it to be the same.

The cumulative radiator function has the form: 
{\small\begin{align}
    \eta_{\rm tot}&=-\eta(\mu_{H},\mu_{F})+\eta(\mu_J,\mu_{cs})\,,\\
    K_{\rm tot}&=-\frac{1}{2}K(\mu_J,\mu_{cs})+K(\mu_H,\mu_{F})-\frac{1}{2}\eta(\mu_J,\mu_{cs})\text{ln}\Big(\frac{\mu_{cs}^2}{Q^2R^2}\Big)+\eta(\mu_{H},\mu_{F})\text{ln}\Big(\frac{\mu_{F}^2}{Q^2}\Big)+\omega_{B}\,,\\
    \mathcal{R}\Big(z,QR;\mu_H^2,\mu_J^2,\mu_F^2,\mu_{cs}^2\Big)&=(1-z)^{\eta_{\rm tot}}\frac{\text{exp}\Big(K_{\rm tot}-\gamma_E\eta_{\rm tot}\Big)}{\Gamma(1+\eta_{\rm tot})}\mathcal{R}^{\rm NGL}(\mu_{J}^2,\mu_{cs}^2)\,.
  \end{align}  }
This defines the radiator as the cumulative distribution, which we then differentiate \emph{after} scale setting. $\gamma_E$ is the Euler-Gamma constant, $\Gamma$ is the gamma function, and the functions $K$ and $\eta$ are integrals over the cusp-anomalous dimension $\Gamma_{\text{cusp}}$ (given to two loops in ref. \cite{Korchemsky:1987wg}), while $\omega_{B}$ is the integral over the non-cusp components of the beam function's anomalous dimension:
\begin{align}
K(\mu_f,\mu_i)&=\int\displaylimits_{\mu_{i}^2}^{\mu^2_f}\frac{d\mu^2}{\mu^2}\Big\{\Gamma_\text{cusp}(\mu^2)\text{ln}\Big(\frac{\mu^2 }{\mu_{i}^2}\Big)\Big\}\,,\qquad\eta(\mu_f,\mu_i)=\int\displaylimits_{\mu_{i}^2}^{\mu^2_f}\frac{d\mu^2}{\mu^2}\Gamma_\text{cusp}(\mu^2)\,,\qquad
\omega_{B}=\int\displaylimits_{\mu_{F}^2}^{\mu^2_H}\frac{d\mu^2}{\mu^2}\gamma_{B}(\mu^2)\,,\\
\Gamma_{\text{cusp}}(\mu^2)&=C_F\frac{\alpha_s(\mu^2)}{\pi}\Bigg(1+\frac{\alpha_s(\mu^2)}{4\pi}\Big(C_A\Big(\frac{67}{9}-\frac{\pi^2}{4}\Big)-\frac{10}{9}n_f\Big)\Bigg)+...,\\
\gamma_{B}(\mu^2)&=C_F\frac{3\alpha_s(\mu^2)}{2\pi}+...\,.
\end{align}
Where we give the anomalous dimensions explicitly for a quark jet. We also have the non-global contribution, which we take as given by the hemisphere distribution for the anti-$k_T$ type algorithms given here:
{\small\begin{align}
    \mathcal{R}^{\rm NGL}(\mu_f^2,\mu_i^2)&=\text{exp}\Big(-C_F C_A\frac{\pi^2}{3}t^2 \frac{1+0.85C_A t^2}{1+(0.86 C_A t)^{1.33}}\Big),\qquad t = \frac{1}{\beta_0}\text{ln}\frac{\alpha_s(\mu_i^2)}{\alpha_s(\mu_f^2)}\,,
\end{align}}
where $\beta_0=\frac{11}{3}C_A-\frac{2}{3}n_f$. We use the two-loop running of the strong coupling constant with $\alpha_s(M_Z^2)=0.1187$, and regulate the Landau pole via the prescription:
\begin{align}
    \mu_{F}^2&\sim Q^2(1-z_{\rm jet})+z_{\rm jet}^2m^2\,,\qquad \mu_{cs}^2\sim Q^2R^2(1-z_{\rm jet})^2+z_{\rm jet}^2m^2\,,\qquad
    m = 0.5~\text{GeV}\,.
\end{align}

 The final cross-section we plot in Fig.~\ref{fig:theory} is given by:
\begin{align}
    \frac{d\sigma}{dz_{\rm jet}}&=\int\displaylimits_{Q_{0}^2}^{s}dQ^2\int\displaylimits_{0}^{1}dx_B\,\Theta\Big(\frac{1-x_B}{x_B}Q^2-Q^2R^2\Big)\Theta\Big(1-\frac{Q^2}{x_Bs}\Big)\frac{d\sigma}{dx_Bdz_{\rm jet}dQ^2}\,.
\end{align}
Where we express the DIS born cross-section $\sigma_0$ in terms of $s$, $Q^2$ and $x_B$. $Q_{0}$ is the minimum hard momentum transfer probed in the measurement. The $\Theta$ function in the integral just states that the invariant mass of the jet must be less than the total invariant mass of the hadronic final state, which is given by $(q+P_h)^2=\frac{1-x_B}{x_B}Q^2$, up to ``target mass corrections.'' We note that we have very little sensitivity to the constraint on $QR$ towards the endpoint $x_B\rightarrow 1$.
    
\FloatBarrier

%
\section{Projections~\label{sec:appA}}

As shown by the fast-detector simulations presented in in Ref.~\cite{Arratia:2020nxw}, the current specifications for future EIC detectors yield a jet-energy resolution that ranges from 15$\%$ to 7$\%$ for jet energy from 15~GeV to 100~GeV, for anti-$k_T$ jets reconstructed in the laboratory frame based on particle-flow objects as implemented in \textsc{Delphes}~\cite{deFavereau:2013fsa}. We estimate a similar performance for jets reconstructed in the Breit frame with the Centauro algorithm. 

\begin{figure}[h!]
  \centerline{\includegraphics[width = 0.42 \textwidth]{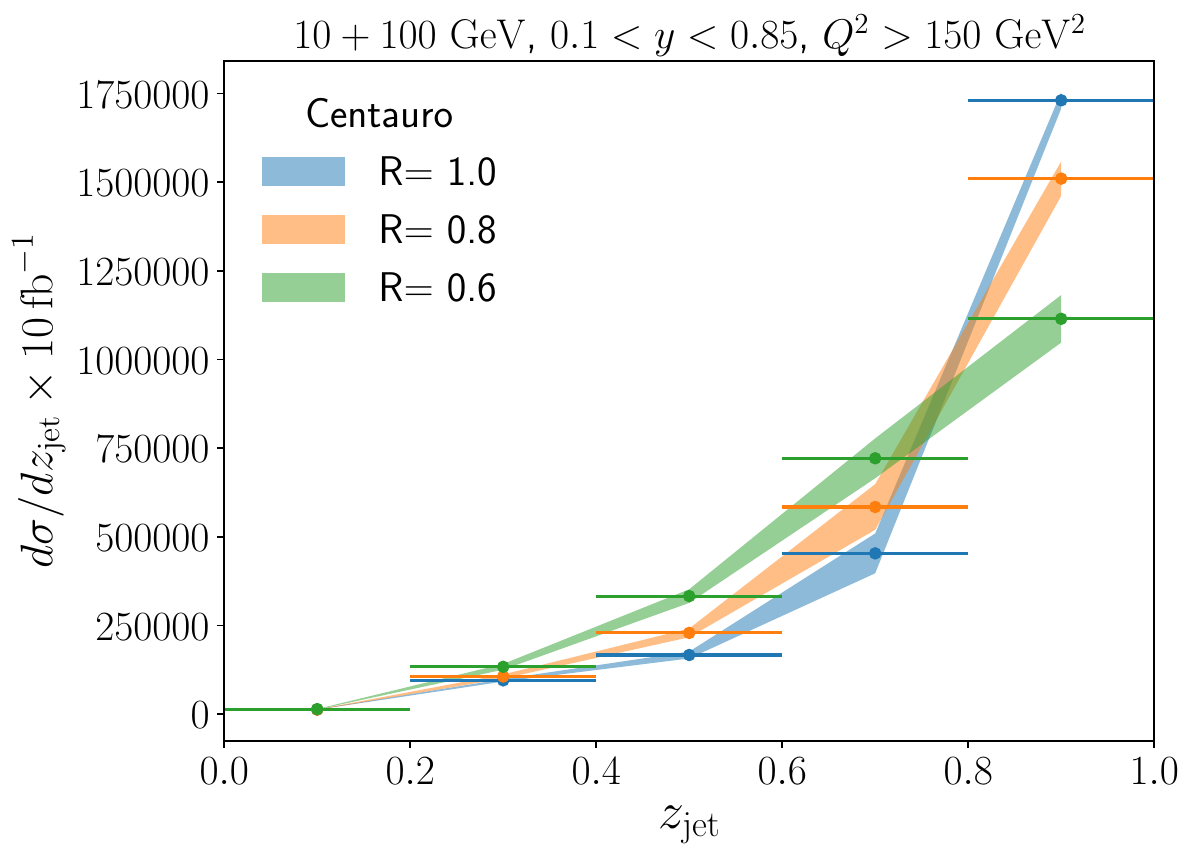}}
  \caption{EIC projecion of the jet-energy spectrum for Centauro jets with different $R$ parameters, assuming an integrated luminosity of 10 fb$^{-1}$ and a jet-energy-scale uncertainty of $\pm2\%$.}
  \label{fig:projections}
\end{figure}

In Figure~\ref{fig:projections} we show a projection of statistical and systematic uncertainties of the jet-energy spectrum for Centauro jets at the future EIC. With an integrated luminosity of 10 fb$^{-1}$, the statistical uncertainty is expected to be negligible. Previous measurements of jets in DIS at HERA suggest that the dominant source of uncertainty will be associated with the jet-energy-scale (JES) calibration, which reached 1$\%$ for the HERA experiments~\cite{Newman:2013ada}. We estimate the JES uncertainty will be worse for the jet-energy spectrum measurement we propose, as the cross-calibration based on electron-jet balance in events close to Born kinematics will not be available (as for this case it represents the signal channel). We thus estimate a JES uncertainty of 2$\%$, which we consider a conservative estimate. The resulting correlated uncertainty on the jet-energy spectrum is shown as a band in Figure~\ref{fig:projections}. 

The bin widths presented in Figure~\ref{fig:projections} are chosen to obtain a controllable unfolding procedure, as  informed by our simulation studies. To obtain a finer binning, an improved calorimeter resolution over current specifications would be required. Alternatively, the jet-energy could be defined with charged-particles only, for which the jet-energy resolution would be better than 1$\%$. Other techniques such as a neutral-hadron-veto could also help to improve the energy resolution~\cite{Page:2019gbf}. We leave dedicated studies to explore these possibilities for future work.

\end{document}